\documentclass[letterpaper]{article} 
\usepackage{aaai2026}  
\usepackage{times}  
\usepackage{helvet}  
\usepackage{courier}  
\usepackage[hyphens]{url}  
\usepackage{graphicx} 
\usepackage{natbib}  
\usepackage{caption} 
\usepackage{algorithm}
\usepackage{algorithmic}

\usepackage{newfloat}
\usepackage{listings}
\DeclareCaptionStyle{ruled}{labelfont=normalfont,labelsep=colon,strut=off} 
\floatstyle{ruled}
\newfloat{listing}{tb}{lst}{}
\floatname{listing}{Listing}
\nocopyright

\title{A Modular LLM-Agent System\\ for Transparent Multi-Parameter Weather Interpretation}
\author{
    Daniil Sukhorukov\textsuperscript{\rm 1, \rm 2},
    Andrei Zakharov\textsuperscript{\rm 1},
    Nikita Glazkov\textsuperscript{\rm 1, \rm 3},
    Katsiaryna Yanchanka\textsuperscript{\rm 4},
    Vladimir Kirilin\textsuperscript{\rm 4},
    Maxim Dubovitsky\textsuperscript{\rm 4},
    Roman Sultimov\textsuperscript{\rm 5},
    Yuri Maksimov\textsuperscript{\rm 6},
    Ilya Makarov\textsuperscript{\rm 1, \rm 7, \rm 8}
}
\affiliations{
    \textsuperscript{\rm 1}AI Research Institute\quad
    \textsuperscript{\rm 2}ITMO University \quad
    \textsuperscript{\rm 3}NUST MISIS\quad
    \textsuperscript{\rm 4}Independent Researcher \quad
    \textsuperscript{\rm 5}AI Center, Lomonosov Moscow State University \quad
    \textsuperscript{\rm 6}LLC Interdata \quad
    \textsuperscript{\rm 7}ISP RAS \quad
    \textsuperscript{\rm 8} Research Center of the Artificial Intelligence Institute, Innopolis University\quad
}

\usepackage{bibentry}

\begin{document}

\maketitle

\begin{abstract}
Weather forecasting is not only a predictive task but an interpretive scientific process requiring explanation, contextualization, and hypothesis generation. This paper introduces AI-Meteorologist, an explainable LLM-agent framework that converts raw numerical forecasts into scientifically grounded narrative reports with transparent reasoning steps. Unlike conventional forecast outputs presented as dense tables or unstructured time series, our system performs agent-based analysis across multiple meteorological variables, integrates historical climatological context, and generates structured explanations that identify weather fronts, anomalies, and localized dynamics. The architecture relies entirely on in-context prompting, without fine-tuning, demonstrating that interpretability can be achieved through reasoning rather than parameter updates. Through case studies on multi-location forecast data, we show how AI-Meteorologist not only communicates weather events but also reveals the underlying atmospheric drivers, offering a pathway toward AI systems that augment human meteorological expertise and support scientific discovery in climate analytics.
\end{abstract}

\begin{links}
    \link{Code}{https://anonymous.4open.science/r/ai-meteorologist/}
\end{links}

\section{Introduction}

Large Language Models (LLMs) are increasingly being applied in meteorology, showing promise in tasks ranging from climate question-answering to tailored forecast advice \cite{em:86}. However, translating raw meteorological data into actionable insights presents significant challenges. Traditional approaches often neglect to convert numerical variables into interpretable narratives of events and their consequences \cite{em:86}. While existing systems focus on extracting insights from textual corpora or generating brief recommendations, they fail to address the end-to-end processing of structured numerical forecast data with rich contextual reasoning \cite{DBLP:journals/kbs/PelaezRodriguezPMPCGS24}. We identify two main challenges: 

\textbf{Meteorological Knowledge Gap:} Most LLMs lack understanding of meteorological contexts, including parameter relationships, atmospheric processes, and regional climatology. Without essential background information, LLMs cannot provide a meaningful analysis of forecast data.

\textbf{Complexity of Weather Data:} Weather forecasts involve multiday, high-frequency predictions across numerous parameters. The LLM must understand temporal dependencies and parameter interactions for accurate interpretation.

In response to these challenges, we propose AI-Meteorologist, a modular agent-based system that translates numerical weather forecasts into structured natural language reports with visualizations. The system employs multiple specialized LLM agents to process structured forecast data, reason about emerging weather patterns, and generate human-readable, explainable reports. By decomposing the task into specialized steps, AI-Meteorologist enables interpretation of detailed forecasts while addressing the scale of the data and the need for trustworthy explanations.

\section{Related Work}

\paragraph{LLM-agent systems for scientific reasoning.} Recent work has demonstrated the emergence of LLM-agent systems as a new paradigm for scientific reasoning. Early systems such as AutoGPT \cite{autogpt} and MetaGPT \cite{hong2023metagpt} introduced autonomous multi-agent collaboration frameworks, while AutoGen \cite{wu2024autogen} formalized conversational agent orchestration for complex analytical tasks. In scientific domains, ChemCrow \cite{m2024augmenting} and AI Scientist \cite{lu2024ai} demonstrated how LLM agents can autonomously design experiments, retrieve literature, and validate hypotheses. These systems collectively illustrate the growing shift from static language models to interactive, tool-augmented scientific agents capable of structured reasoning and knowledge generation.

\paragraph{LLMs in meteorology and weather forecasting.}
Recent work has begun leveraging LLMs to translate structured meteorological data into human-interpretable weather narratives and interactive tools. For example, ECMWF’s DestinE chatbot \cite{ecmwf_destine_chatbot_2025} to make high-resolution weather and climate data accessible via conversational interfaces, CLLMate \cite{li2024cllmate} to enable event-based forecasting of weather and climate phenomena in text form, and GPT-based forecasting \cite{franch2025gptcast} to perform precipitation nowcasting by tokenizing radar imagery and generating ensemble forecasts via language-model driven frameworks. More recent works have explored LLM-agent frameworks that integrate iterative querying, reflection, and domain-specific validation to improve the scientific accuracy and robustness of weather reports \cite{varambally2025zephyrus}.

\section{System Description}

The system employs a modular agent-based LLM architecture designed for the automated generation of explainable weather forecast reports from structured prediction data. Operating as a pipeline of specialized agents, each component handles the distinct aspects of transforming raw meteorological input into human-readable bulletins. Implementation leverages the GPT-4o API and processes high-resolution time-series data retrieved from the OpenWeather One Call 2.5 API, providing hourly forecasts up to five days ahead.

\textbf{Data Acquisition and Preprocessing}. The pipeline begins with data acquisition for a selected location, retrieving forecast time series via the OpenWeather API. This includes more than ten meteorological parameters, including temperature, humidity, wind, precipitation, and visibility at a resolution of one hour. The system simultaneously gathers contextual information, comprising geographic metadata from OpenStreetMap (such as region type and urban/rural classification) and climatological norms from the Meteostat API, specifically 20-year monthly averages of temperature and precipitation. All collected data are serialized into a structured EXTERNAL INFO block and propagated downstream to the reasoning agent.

\textbf{Meteorologist Agent}. The core analytical component generates a structured output containing four key fields: a natural language summary outlining expected weather evolution; a causal explanation grounded in meteorological reasoning; a self-assessed confidence level; and warnings describing potentially hazardous or anomalous conditions detected through deviation from climatology.

\textbf{Writer Agent}. Building on the meteorologist's output, the Writer agent creates the report's textual structure: a title (location/parameters), an introduction synthesizing outlook with summary, proof, and warnings, and a weather\_params block with key variable descriptions. The text emulates official bulletins, optionally incorporating user preferences.

\textbf{Illustrator Agent}. Based on the selected parameters, this agent generates Python code using matplotlib to produce plots of forecast data such as temperature, wind, and precipitation, ensuring that visual representations align with the textual narrative.

\textbf{Report Compilation and Output}. The final report integrates all content blocks into a structured PDF document, comprising a general synopsis, detailed forecast summary and reasoning, visual plots, and optional warnings with climatological comparisons. This comprehensive output provides users with both textual explanations and visual representations of the forecast data.

\paragraph{Sample Prompt:} 

\textit{You are a professional meteorologist tasked with interpreting a time series of weather data for a specific location. Your goal is to write a clear, concise, and scientifically grounded summary of the weather conditions over the given period.}

\textit{You must: - Describe the evolution of key meteorological parameters such as temperature, wind, precipitation, humidity, and pressure over time. - Identify and explain significant *weather phenomena*, including heat waves, cold fronts, thunderstorms, or abrupt changes in conditions. - Highlight temporal *trends*, such as warming, cooling, increasing cloudiness, or shifting wind directions. - Provide *interpretive insights*, not just numerical repetitions — describe what the numbers imply about the atmosphere. - If provided, use climatological normals or geographic context to comment on anomalies (e.g., “unusually cold for this region”). - Avoid simply listing values or copying table entries. Instead, synthesize the data into meaningful meteorological observations. - Use CLIMATOLOGY INFO from input data for describing anomaly events.}

\textit{Use formal meteorological language but remain understandable to non-specialists. Structure your output in coherent paragraphs, typically 4–6 sentences, covering the full time period.}

\textit{You are NOT generating a forecast. You are summarizing and interpreting **past or present** observed or modeled weather data.}

\section{Results}

\begin{figure}[t]
\centering
\includegraphics[width=0.45\textwidth]{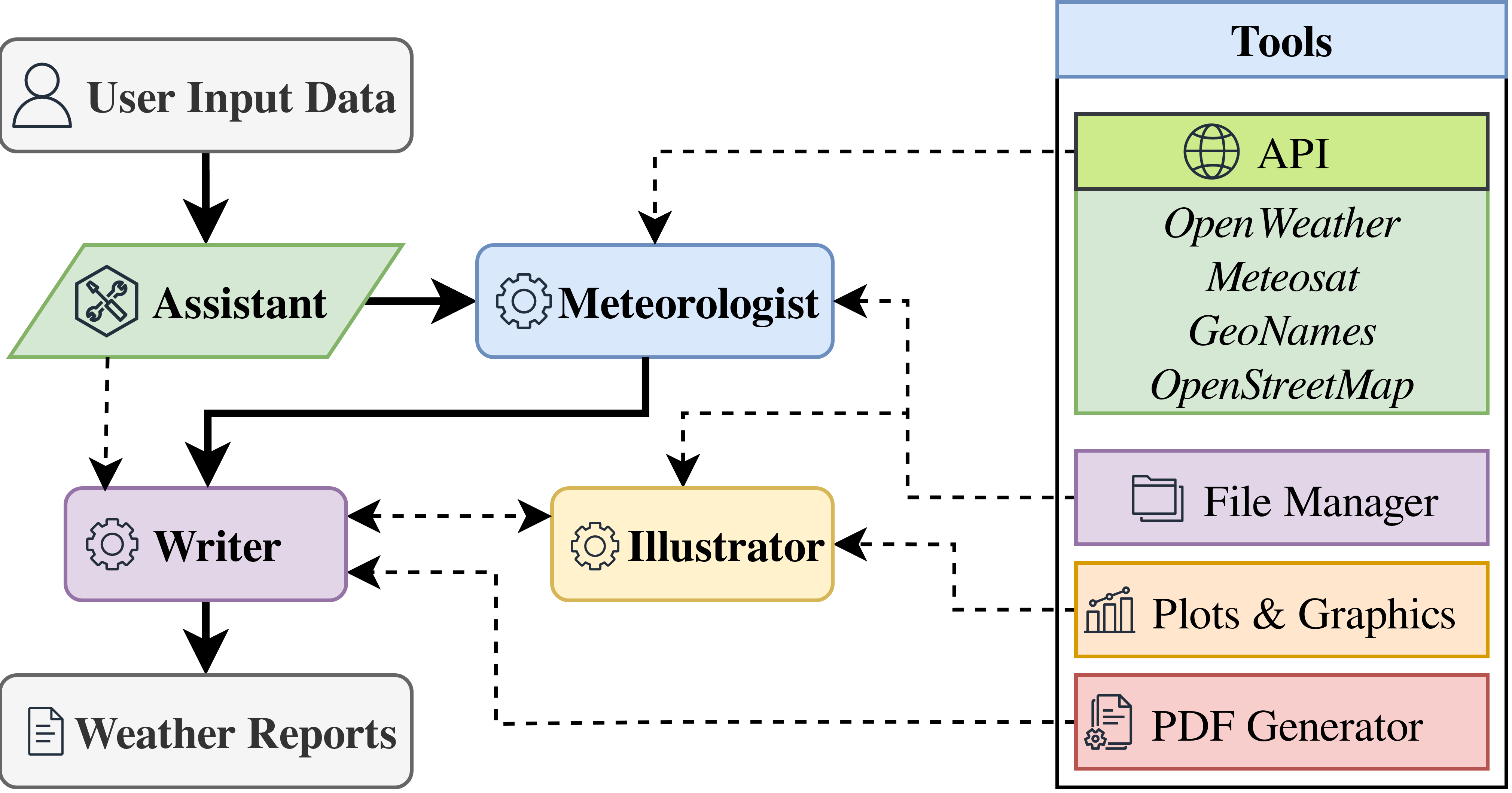}\\
\vspace{10 pt}
\includegraphics[width=0.47\textwidth]{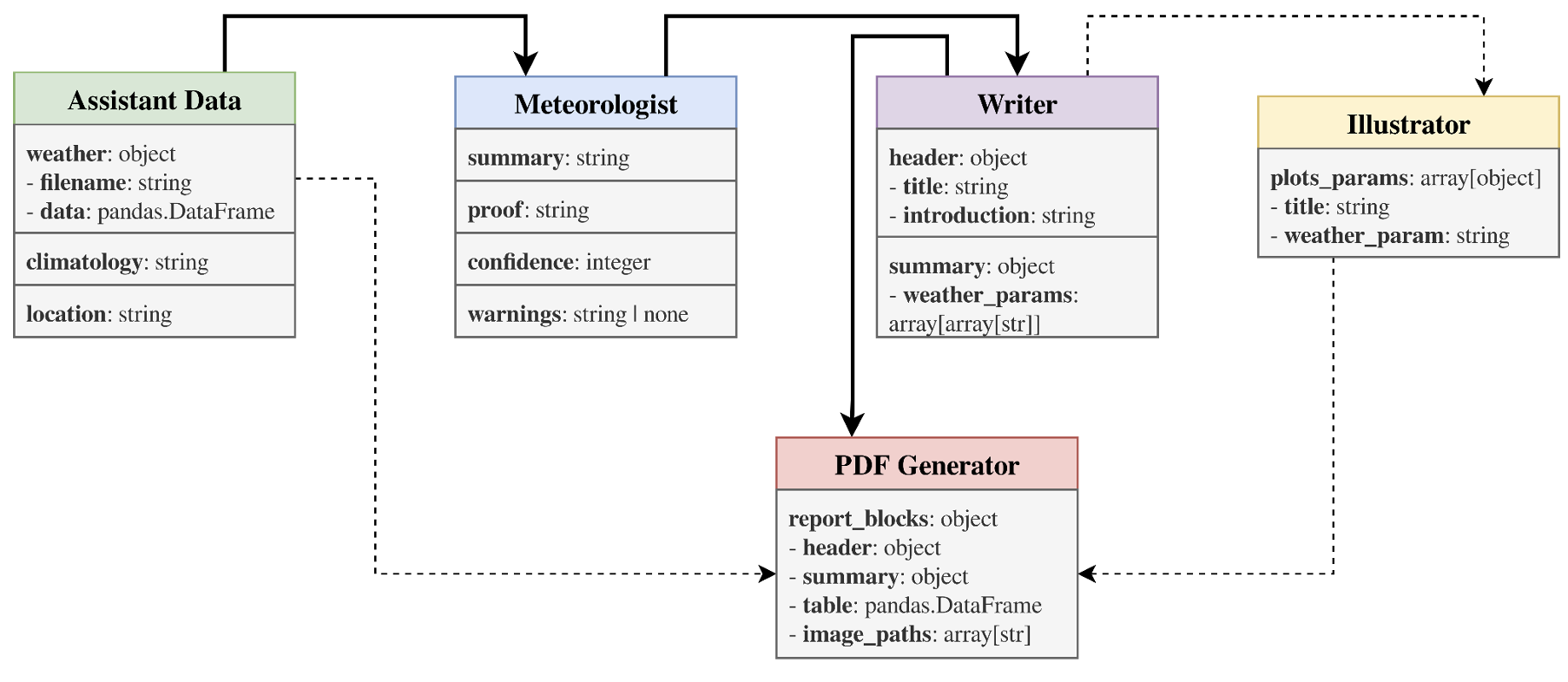}
\caption{Interaction of individual system modules for generating weather reports. Core workflow is supported by weather APIs, file management, visualization capabilities, and end-to-end report generation.}
\label{architecture}
\end{figure}

\begin{figure*}[t]
\centerline{\includegraphics[width=0.95\textwidth]{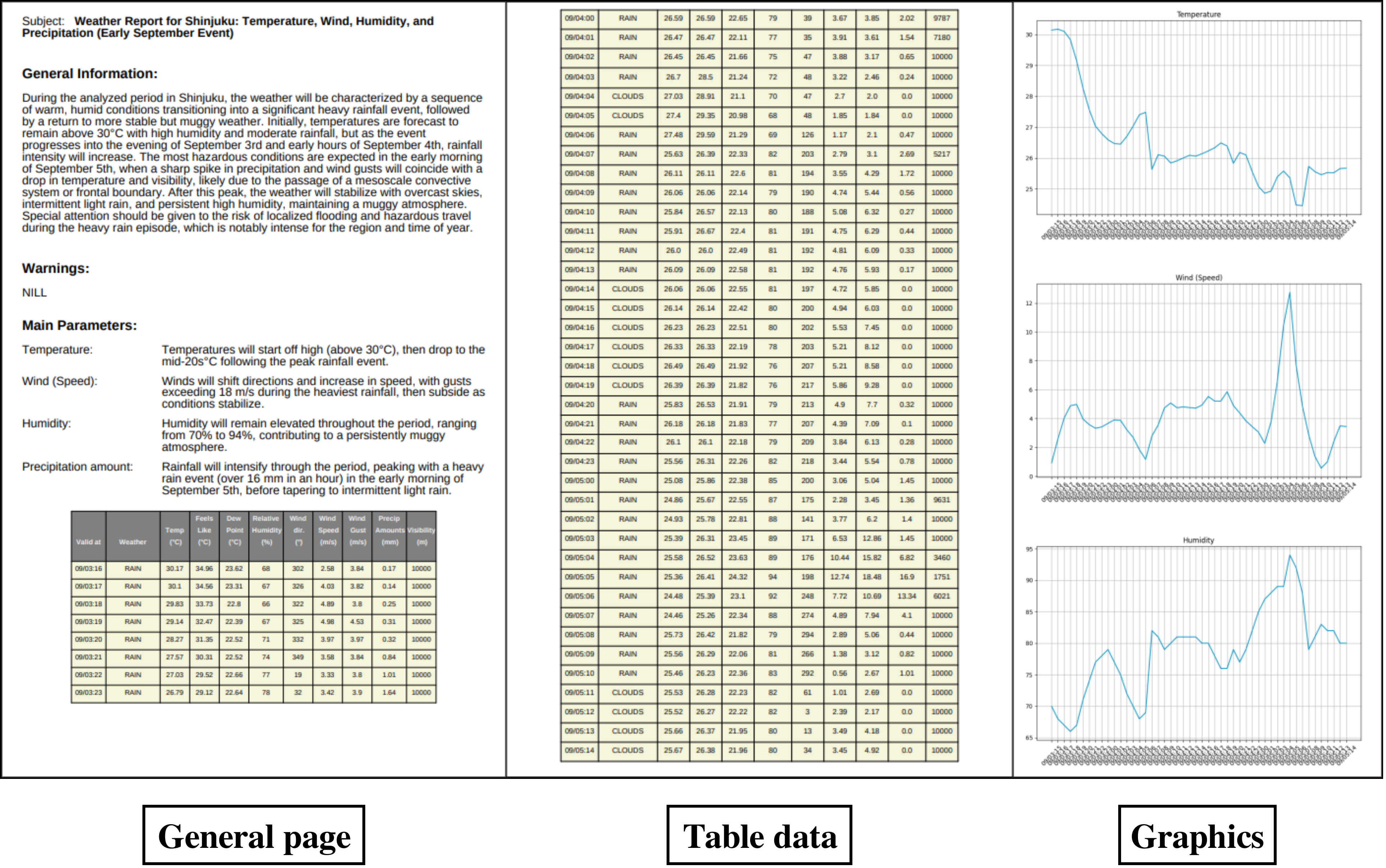}}
\caption{Example of a report generated by the system with meteorological insights, detailed multi-parameter forecasts, and time-series visualizations of key weather variables.}
\label{docements}
\end{figure*}

We evaluate the AI-Meteorologist system on real-time multi-parameter forecast data across coastal, inland, and high-altitude locations to assess its ability not only to generate reports, but to conduct interpretable scientific reasoning. Figure \ref{architecture} illustrates the interaction between the agent modules during inference, and Figure \ref{docements} presents an example forecast report automatically generated without fine-tuning.


Across all test cases, the system consistently identified key meteorological phenomena through explainable diagnostic reasoning steps. For a coastal location (Synoptic Front Detection) the agent detected an approaching cold front by jointly analyzing pressure gradients, wind direction rotation, and a rapid decrease in temperature. Rather than merely reporting numerical changes, the system generated a scientifically grounded explanation describing how frontal boundaries form due to interactions between maritime and continental air masses. It further contextualized the event using local climatological baselines and automatically generated a confidence score indicating forecast reliability. This demonstrates the system’s ability to move beyond narrative reporting toward hypothesis-driven meteorological interpretation. For an inland region (Anomaly and Hazard Reasoning), the system identified anomalous precipitation by comparing short-term forecasts with long-term climatological percentiles. The agent explained the deviation from expected seasonal patterns and issued a data-driven risk statement regarding potential flooding. Importantly, the explanation included its reasoning chain, referencing historical anomalies and presenting the conclusion as a structured argument rather than a black-box output.

These results demonstrate that the AI-Meteorologist is not merely a language interface, but a reasoning-capable scientific assistant. By combining hierarchical interpretation, semantic keyword extraction, and climatological knowledge integration, the system provides transparent and reproducible insights—fulfilling the goals of explainable AI for advancing scientific understanding of weather systems.

\section{Discussion}
The AI-Meteorologist system demonstrates several key advances in automated weather forecasting and explanation. First, it addresses the critical challenge of translating dense numerical meteorological data into human-interpretable narratives without requiring domain-specific fine-tuning of the underlying LLM. This approach significantly reduces the computational resources and expertise typically needed for specialized AI-weather systems \cite{zhang2025streamlining}.

Second, the system's modular agent-based architecture provides flexibility and extensibility, enabling incremental improvements without disrupting the overall workflow. Third, the integration of multiple data sources enables comprehensive contextual reasoning that references both immediate weather patterns and the long-term climatic context.

However, several limitations warrant consideration. The performance of the system depends on the quality and coverage of the underlying APIs and data sources. Additionally, while the current implementation demonstrates strong capabilities in generating textual explanations, the quantitative evaluation of forecast accuracy and explanation quality remains an area for future work.

\section*{Conclusion}
We presented AI-Meteorologist, a modular LLM-agent system designed to generate explainable weather forecast reports from structured meteorological data. Our approach integrates time-series reasoning, climatological context, and geospatial metadata to produce high-quality bulletins that go beyond numerical repetition, offering interpretability, causal justifications, and warnings about anomalous conditions.

Our demonstration illustrates that even general-purpose LLMs, such as GPT-4o, can reason over dense hourly forecast tables when provided with properly serialized inputs and structured prompts. The modularity of our architecture allows for extensibility, including multilingual support, user-driven customization of the report structure, and incorporation of additional forecast sources.

Future directions include comparative evaluation of generated reports against human-authored bulletins, integration with probabilistic ensemble data, and exploration of self-refinement loops for improved robustness. By bridging the gap between raw meteorological data and human-interpretable explanations, AI-Meteorologist represents a significant step toward more accessible and trustworthy automated weather forecasting systems.

\section*{Acknowledgments and Disclosure of Funding}
The work of I. Makarov was supported by the Ministry of Economic Development of the RF (agreement No. 139-10-2025-034 dd. 19.06.2025, IGK 000000C313925P4D0002)

\bibliography{aaai2026}

@book{em:86,
  editor  = "Engelmore, Robert and Morgan, Anthony",
  title   = "Blackboard Systems",
  year    = 1986,
  address = "Reading, Mass.",
  publisher = "Addison-Wesley",
}

@article{zhang2025streamlining,
  title={Streamlining geoscience data analysis with an LLM-driven workflow},
  author={Zhang, Jiyin and Clairmont, Cory and Que, Xiang and Li, Wenjia and Chen, Weilin and Li, Chenhao and Ma, Xiaogang},
  journal={Applied Computing and Geosciences},
  volume={25},
  pages={100218},
  year={2025},
  publisher={Elsevier}
}

@article{DBLP:journals/kbs/PelaezRodriguezPMPCGS24,
  author       = {C{\'{e}}sar Pel{\'{a}}ez{-}Rodr{\'{\i}}guez and
                  Jorge P{\'{e}}rez{-}Aracil and
                  Cosmin Madalin Marina and
                  Luis Prieto{-}Godino and
                  Carlos Casanova{-}Mateo and
                  Pedro Antonio Guti{\'{e}}rrez and
                  Sancho Salcedo{-}Sanz},
  title        = {A general explicable forecasting framework for weather events based
                  on ordinal classification and inductive rules combined with fuzzy
                  logic},
  journal      = {Knowl. Based Syst.},
  volume       = {291},
  pages        = {111556},
  year         = {2024},
  url          = {https://doi.org/10.1016/j.knosys.2024.111556},
  doi          = {10.1016/J.KNOSYS.2024.111556},
  bibsource    = {dblp computer science bibliography, https://dblp.org}
}

@article{li2024cllmate,
  title={CLLMate: A Multimodal Benchmark for Weather and Climate Events Forecasting},
  author={Li, Haobo and Wang, Zhaowei and Wang, Jiachen and Wang, YueYa and Lau, Alexis Kai Hon and Qu, Huamin},
  journal={arXiv preprint arXiv:2409.19058},
  year={2024}
}

@article{lu2024ai,
  title={The ai scientist: Towards fully automated open-ended scientific discovery},
  author={Lu, Chris and Lu, Cong and Lange, Robert Tjarko and Foerster, Jakob and Clune, Jeff and Ha, David},
  journal={arXiv preprint arXiv:2408.06292},
  year={2024}
}

@article{m2024augmenting,
  title={Augmenting large language models with chemistry tools},
  author={M. Bran, Andres and Cox, Sam and Schilter, Oliver and Baldassari, Carlo and White, Andrew D and Schwaller, Philippe},
  journal={Nature Machine Intelligence},
  volume={6},
  number={5},
  pages={525--535},
  year={2024},
  publisher={Nature Publishing Group UK London}
}

@inproceedings{hong2023metagpt,
  title={MetaGPT: Meta programming for a multi-agent collaborative framework},
  author={Hong, Sirui and Zhuge, Mingchen and Chen, Jonathan and Zheng, Xiawu and Cheng, Yuheng and Wang, Jinlin and Zhang, Ceyao and Wang, Zili and Yau, Steven Ka Shing and Lin, Zijuan and others},
  booktitle={The Twelfth International Conference on Learning Representations},
  year={2023}
}

@inproceedings{wu2024autogen,
  title={Autogen: Enabling next-gen LLM applications via multi-agent conversations},
  author={Wu, Qingyun and Bansal, Gagan and Zhang, Jieyu and Wu, Yiran and Li, Beibin and Zhu, Erkang and Jiang, Li and Zhang, Xiaoyun and Zhang, Shaokun and Liu, Jiale and others},
  booktitle={First Conference on Language Modeling},
  year={2024}
}

@misc{autogpt,
  author       = {Significant-Gravitas},
  title        = {AutoGPT},
  year         = {2023},
  howpublished = {\url{https://github.com/Significant-Gravitas/AutoGPT}},
  note         = {Version v0.6.33, accessed: 2025-10-22}
}

@article{franch2025gptcast,
  title={GPTCast: a weather language model for precipitation nowcasting},
  author={Franch, Gabriele and Tomasi, Elena and Wanjari, Rishabh and Poli, Virginia and Cardinali, Chiara and Alberoni, Pier Paolo and Cristoforetti, Marco},
  journal={Geoscientific Model Development},
  volume={18},
  number={16},
  pages={5351--5371},
  year={2025},
  publisher={Copernicus Publications G{\"o}ttingen, Germany}
}

@article{varambally2025zephyrus,
  title={Zephyrus: An Agentic Framework for Weather Science},
  author={Varambally, Sumanth and Fisher, Marshall and Thakker, Jas and Chen, Yiwei and Xia, Zhirui and Jafari, Yasaman and Niu, Ruijia and Jain, Manas and Manivannan, Veeramakali Vignesh and Novack, Zachary and others},
  journal={arXiv preprint arXiv:2510.04017},
  year={2025}
}

@misc{ecmwf_destine_chatbot_2025,
  author       = {ECMWF},
  title        = {Development Seed to create a climate and weather chatbot in DestinE},
  year         = {2025},
  month        = {April},
  day          = {2},
  howpublished = {“News” article on the Destination Earth website},
  note         = {\url{https://destine.ecmwf.int/news/development-seed-to-create-a-climate-and-weather-chatbot-in-destine/?utm_source=chatgpt.com} (accessed: 2025-10-22)}
}


\onecolumn
\section*{Supplementary Material}

\begin{figure*}[h]
\centering
\includegraphics[width=0.95\textwidth]{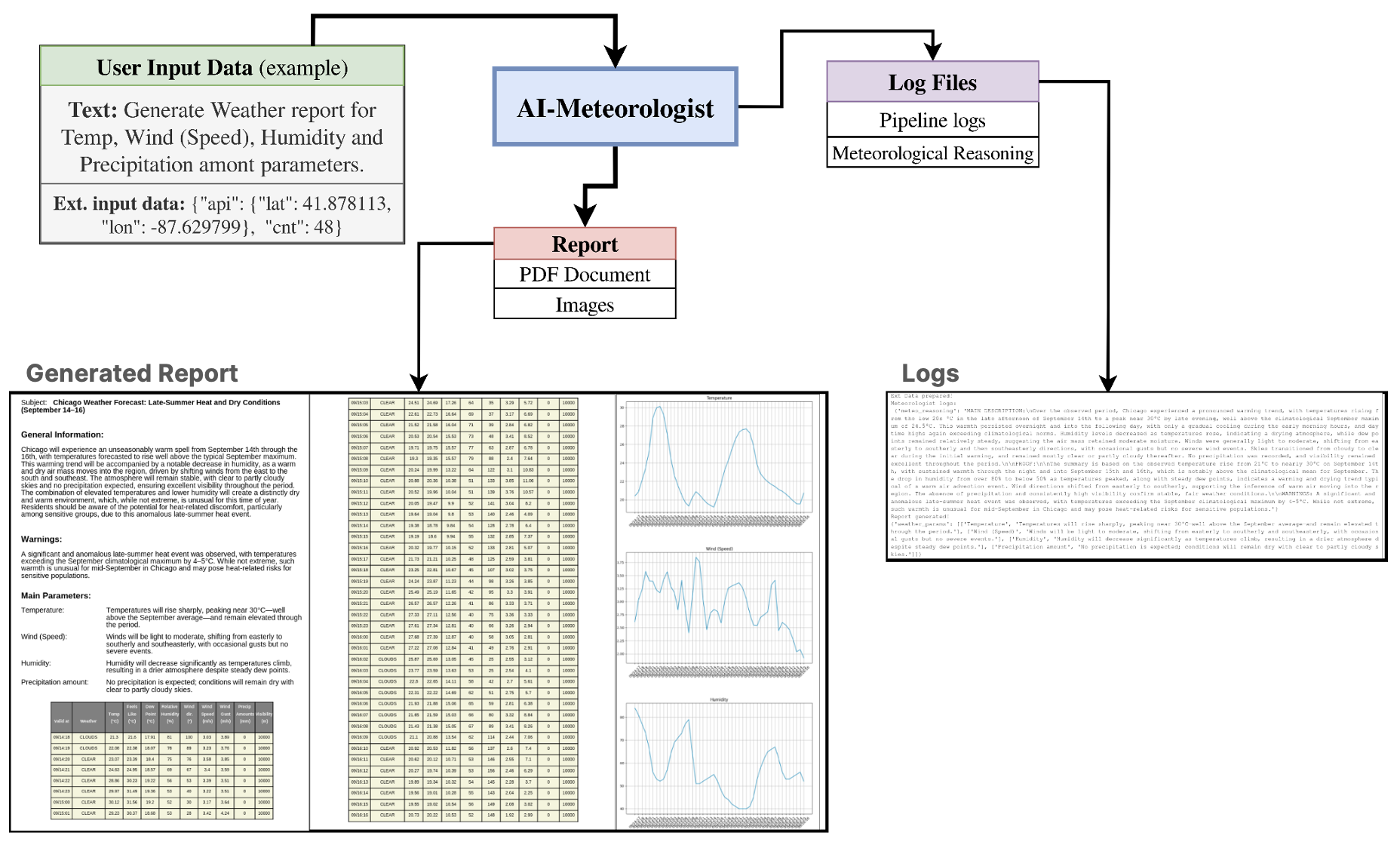}
\caption{Schematic overview of the weather report generation process.}
\label{example}
\end{figure*}

\end{document}